\documentclass[aps,prl,twocolumn,groupedaddress,showpacs]{revtex4}

\usepackage{graphicx}

\begin{document}


\title{Brownian motion of fractal particles: L\'evy flights from white noise }


\author{Kiran M. Kolwankar}
\affiliation{Max Planck Institute for Mathematics in the Sciences,
Inselstrasse 22,
D - 04103 Leipzig, Germany}


\date{\today}

\begin{abstract}
We generalise the Langevin equation with Gaussian white noise by replacing the
velocity term by a local fractional derivative. 
The solution of this equation is a L\'evy process.
We further consider the
Brownian motion of a fractal particle, for example, a colloidal
aggregate or a biological molecule and argue that
it leads to a L\'evy flight. This effect can also be described 
using the local fractional Langevin equation.
The implications of this development to other complex data
series are discussed.
\end{abstract}


\maketitle

\section{}
Exactly 100 years ago Einstein~\cite{Einstein1905}
 proposed the theory of Brownian
motion which marked the birth of the field of stochastic processes
and has worked as a guiding principle in modelling of many
fundamental as well as applied phenomena in
physical, chemical and biological sciences (see 
reviews~\cite{Frey2005,Hanggi2005} and references therein
 which document 
its impact).
Present interest in complex high dimensional systems,
in wide ranging fields, whose 
dynamics may not be completely known has given rise to 
the study of many irregular processes. 
Experience shows that, in spite of
the lack of understanding of the laws governing the phenomena,
they can be successfully modelled using stochastic processes.
As a result new processes are being introduced~\cite{Eliazar2005}.
In the past, the Gaussian process used to be the main tool employed to this end.
But recently, owing to the frequent occurrence of 
anomalous diffusions~\cite{Bouchaud1990}
(where $<x^2> \propto t^{\alpha}$, $\alpha \neq 1$),
two main
generalisations of Gaussian process, viz., 
L\'evy processes~\cite{Shlesinger1995} and 
fractional Brownian motion (fBm)~\cite{Mandelbrot1968}, 
have turned out to be of
importance. These processes respectively relax two
important assumptions in the central limit theorem that of
finite variance and independence. A process whose
second moment diverges but the first moment is finite
falls into the domain of attraction the L\'evy process
of index $\mu$, with $1<\mu<2$, where $\mu$
characterises the power law tail of the probability
distribution function and the one with even the
first moment infinite corresponds to the 
L\'evy process with $0<\mu<1$.
It is now known that many economical time series
are better modelled by a truncated L\'evy flight~\cite{Mantegna2000}. Also in
biological physics, modern high speed imaging techniques have
uncovered many interesting anomalous diffusive behaviours of
large biological molecules~\cite{Ritchie2005}. 

On the other hand, derivatives and integrals of non-integer 
order~\cite{Samko1993,Podlubny1999} 
have been found to be useful in successfully describing scaling processes.
The realm of applications of such a fractional calculus is fast
expanding with ever new developments rapidly
taking place in the field of statistical and nonlinear physics
over the last few years~\cite{Metzler1999,Hilfer2000,Metzler2000,Zaslavsky2002,
West2003,Metzler2004,Grigorenko2003,Carpinteri1997}.
Fractional derivatives were used in  the
definition of fBm~\cite{Mandelbrot1968,Sebastian1995} 
and the fractional Langevin equation has
been shown to give rise to fBm~\cite{Lutz2001b}. Many researchers have
used the diffusion equation involving fractional derivative
in space which describes a L\'evy process 
and there are
studies wherein the Gaussian white noise 
in the Langevin equation is replaced by a
L\'evy noise~\cite{Gorenflo1999,Brockmann2003,Fogedby1994a,Fogedby1994b,Lutz2001a}. 
This models the anomalous behaviour of the environment or the heat bath.
But there is no
generalisation of the Langevin equation in which one naturally 
obtains a L\'evy process from usual white noise. 
That is the heat bath is normal but the system is irregular and
there is no memory.
It has been shown, however, that the L\'evy flights can be obtained
from continuous time random walks by introduction of an operational
time~\cite{Sokolov2000}.

In this letter we generalise the Langevin equation by incorporating
the local fractional derivatives and show that it leads to
the L\'evy flights from usual white noise. 
We then consider, as an example, the Brownian motion of
rigid irregular particles, a study possibly relevant for aggregates and 
biological molecules. 
We argue that the
fractal nature gives rise to the L\'evy flights. We describe
this phenomena using the local fractional
Langevin equation. We further point out the importance of
this formalism in describing irregular processes which arise
in econophysics.

One way of defining the fractional derivative is through the
so called
Riemann-Liouville fractional derivative~\cite{Samko1993,Podlubny1999}. 
For $q$, the order of the derivative, between zero
and one it is given by:
\begin{eqnarray}
{_aD_{x}^{q}}f(x) = {1\over\Gamma(1-q)}{d\over dx}
\int_a^x{f(y)\over (x-y)^{q}}dy.
\end{eqnarray}
Clearly this is nonlocal and depends on the lower limit $a$. It is
interesting to note that it is this dependence on the parameter, which
led to confusion in different ways of defining a fractional
derivative towards the end of 19th century and hence possibly 
delayed its application, is now becoming important.
The value of $a$ is usually dictated by the 
problem one is investigating.
In some cases it is appropriate to put it equal to zero and in some other it 
is taken to be
$\pm \infty$. In the latter case it is called Weyl derivative. Another
choice of the lower limit is made in the following definition called the
Local fractional derivative (LFD) defined as 
follows~\cite{Kolwankar1996,Kolwankar2001}:
\begin{eqnarray}\label{eq:deflfd}
f^{(q)}(x) = \lim_{x'\rightarrow x}{_{x'}D_{x}^{q}}(f(x)-f(x'))
\;\;\;\; 0<q<1.
\end{eqnarray}
where we introduce the notation that
the superscript in the brackets on the LHS denotes the LFD
of that order.
This definition naturally appears in the local fractional Taylor
expansion~\cite{Kolwankar1996} 
giving it a geometrical interpretation. It should be noted
that the extra limit in the definition of the LFD makes it very different
from other definitions of fractional derivatives and some of its properties
are very different from the non-local versions of fractional derivatives
(see~\cite{Adda2001,Babakhani2002,Krishnamurthy2000,Carpinteri2002,
Carpinteri2004} for more mathematical properties 
and some applications of the LFD).
One important
difference is that though it reduces to the usual derivative when $q=1$,
the LFD is not an analytic function of $q$ for a given
function. For example, if the function is smooth then
the LFD of any order $q$ less than one is zero. In fact, in general
for any continuous nondifferentiable function
there exists a critical order of differentiability 
between zero and one below which
all the derivatives are zero~\footnote{As of now, we do not know
any example of a continuous everywhere but 
nowhere differentiable function for which the LFD 
exist at the critical
order. But here we take a point of view that such functions nevertheless
exist.} and above which they do not exist. 
This critical order of differentiability is equivalent to
the local H\"older exponent
or the local power law exponent~\cite{Kolwankar2001}.
and for
this formalism to yield meaningful results we should work at the
order which is equal to this exponent.
So it can be said that the LFD is a nonanalytic extension of the usual derivative.
This means that the LFD signals an emergence of a new calculus with its new
rules to which one should get accustomed. 
This is much in the same way as in the case of, for
example, the Ito calculus.
It should be noted that the LFD characterises the local scaling
whereas the nonlocal fractional derivatives are useful to study
asymptotic scaling. Owing to these complimentary roles played by
these two versions of the definitions it can not be ruled out that
in some applications
a combination of the two is indispensable.
The limit in eqn~(\ref{eq:deflfd}) which is akin to the 
limit in the renormalization group
transformation, in fact, bestows the LFD a physical interpretation. 
It can be used to relate and study the dynamics of
renormalised quantities. Carpinteri and
Cornetti~\cite{Carpinteri2002} used it to relate renormalised stress and 
strain when the stress
is concentrated on a singular set.

The next step is to consider the equations involving the LFD. The simplest
such equation is
\begin{eqnarray}
f^{(q)}(x) = g(x)
\end{eqnarray}
where $g(x)$ is a known function and $f(x)$ an unknown. Using
the local fractional Taylor expansion~\cite{Kolwankar1996} 
its solution can be written as 
a generalised Riemann sum giving
\begin{eqnarray}\label{eq:soln}
f(x) = \int g(x) d^{q}x = \lim_{N\rightarrow\infty} \sum_{i=0}^{N-1} g(x_i^*)
{(x_{i+1}-x_i)^q\over\Gamma(q+1)}
\end{eqnarray}
where $g(x_i^*)$ is an appropriately chosen point in the interval
$[x_i,x_{i+1}]$. 
Such integrals should be useful for integrating physical quantities
over  a fractal boundary, for example, a current passing through
an interface. The order of the integral, in this case, of course
being equal to the H\"older exponent of the fractal curve. 
For $q<1$, if the function $g(x)$ is, say,  nonnegative in some
interval then the above sum grows in the limit~\footnote{It is 
interesting to note that though such integrals arise naturally
in our formalism as inverse of LFD, Mandelbrot already
in~\cite{Mandelbrot1982} has suggested studying such integrals using nonstandard
analysis which extends the real number system to include infinite
and infinitesimally small number. However, as is made clear in
the following, since we restrict $g(x)$ to two classes of 
physically meaningful functions for which this integral is finite
it obviates the need to use the nonstandard analysis}. Yet,  
two classes of functions $g(x)$ can be identified which will
yield a nontrivial function $f(x)$ as a solution. The first class corresponds
to the functions which have fractal support~\cite{Kolwankar1998}. 
Then if $q=\alpha$ the
dimension of the support of the function the sum above converges 
since only a few terms contribute to the sum giving
rise to a finite solution. This function, called the "devil's
staircase", changes only on
the points of the support of $g(x)$ and is constant everywhere else.
We denote this solution by $P_g(x)$. 
The second class, which is a main focus of this work, 
consists of rapidly oscillating
functions which oscillate around zero in any small interval. 
These oscillations then result in cancellations again giving
rise to a finite solution.
A realisation of the white noise
is one example in this class of functions. This immediately prompts us to
consider a generalisation of the Langevin equation which involves LFD
and $g(x)$ is chosen as white noise.

So we consider a generalisation of the 
Langevin equation~\cite{Risken1989} in the high friction
limit where one neglects the acceleration term
and replaces the first derivative term by the LFD to arrive at
\begin{eqnarray}\label{eq:lengevin}
x^{(\alpha)}(t) = \zeta(t), 
\end{eqnarray}
where $<\zeta(t)>=0$ and $<\zeta(t)\zeta(t')>=\delta(t-t')$
the Dirac delta function.
The solution of the above equation follows from Eq.~(\ref{eq:soln}).
Heuristically it can be seen that
\begin{eqnarray}
<x(t)> &=& t^{\alpha}\lim_{N\rightarrow\infty} 
{N^{-\alpha}\over\Gamma(\alpha+1)} N^{1/2} 
\end{eqnarray}
and therefore the average is zero if $\alpha > 1/2$ and it does not
exist if $\alpha < 1/2$. 
Now we consider the second moment.
\begin{eqnarray}
<x(t_1)x(t_2)> 
&=&  \int_0^{t_1}\int_0^{t_2} \delta(t_1'-t_2')   d^{\alpha}t_1' d^{\alpha}t_2'
\end{eqnarray}
which exists only when $\alpha=1$. 
In order to see this systematically we generalise the concept of
the delta function. The usual Dirac delta function is defined as the
derivative of the Heaviside step function ($\theta(x)=0$ for $x<0$
and 1 for $x > 0$); $\delta(x) = \theta '(x)$. Here two things can
be generalised, first the derivative can be replaced by the LFD
of order $\gamma$ and second, the Heaviside function can be replaced
by a scaling function $\bar{\theta}(x)=x^{\beta}$ for $x>0$ and
0 for $x<0$ and $0<\beta<1$. As a result, our generalised delta
function is defined as $\delta(x;\gamma,\beta)=\bar{\theta}^{(\gamma)}(x-y)
|_{x=y}$. It follows from this definition that $\int \delta(x;\gamma,\beta)
d^{\alpha}x = 0$ for $\alpha > \gamma - \beta$ and $\infty$ when
$\alpha < \gamma - \beta$. Now $\delta(x) = \delta(x;1,0)$ leading to
the above conclusion.
This shows that when $0 < \alpha 1/2$ both the first and the
second moments diverge whereas when $1/2<\alpha<1$ only the
second moment diverges.
This implies that the above process is
in the domain of attraction of
a L\'evy process of index $2\alpha$ for $\alpha < 1$. 
This brings forth some interesting perspective.
The generalisation of the Langevin
equation using non-local fractional operators gives rise to the
fractional Brownian motion whereas the one using the local fractional
operator results in the L\'evy process.

As an application of this formalism we consider the Brownian motion
of a rigid fractal particle. Such  consideration is useful in colloids
as well as biological systems~\cite{Dewey1997}. 
Usually theoretical studies of Brownian motion are restricted to spherical
particles in which case one has, from the Stokes' law, the
formula for frictional force in terms of the viscosity and 
one also assumes that the displacements well separated in time
are not correlated.
Any deviation from the sphericity makes matters 
complicated~\cite{Subramanian1975}.
Berry~\cite{Berry1989} studied velocity of fractal flakes
falling under the gravity. He assumed that the cluster
entrains the air inside it and used the Stokes' law. 
We do not make any such assumption in the following.
Also, there exists a formal theoretical treatment of
the Brownian motion of particles of arbitrary shapes 
using hydrodynamical approach~\cite{Fox1970}.
However it is not valid for fractal particles since
one needs to solve fluctuating hydrodynamical
equations with boundary conditions emanating from the surface
of the particle and in the process one needs to integrate the
normal components of the fluctuating stress tensors over the surface.
This can not be carried out for a fractal particle since the
normal can not be defined on a fractal boundary and integrating 
over a fractal surface would require integrals as in 
equation~(\ref{eq:soln}) necessitating a new approach using the
present calculus with the order of the derivative being the
local H\"older exponent of the surface.
Here we model the irregularity of the suspended particle by
the fractality and first, using purely statistical arguments, 
argue that the Brownian motion of
such irregular particles leads to the L\'evy flights. 
Then we arrive at the same conclusion starting from
above local fractional Langevin equation. The essential
step is to compare the distribution of the
resultant force acting along the center of mass on a 
particle with fractal boundary with the
lateral dimension $D$ to that of a
spherical particle of
diameter $D$. We treat the problem in two dimensions.
First, we consider the ideal situation wherein the
boundary is a mathematical fractal without any lower cutoff and
the surrounding fluid consists of point particles. 
We ignore any lower length scales in the problem.
In a small time interval $\Delta t$, $N$ particles
collide with a spherical Brownian particle whereas for a particle
with a fractal boundary $N^d$ particles, with $d$,
$1<d<2$ being the dimension
of the fractal boundary, will impart their momentum to
the fractal particle. Given the fact that the case with a spherical
particle gives rise to the normal diffusion with finite variance
leads us to the conclusion that the fractal particle,
under the same assumptions, will undergo
diffusion with much larger fluctuations leading it to the basin of
attraction of a L\'evy process. More precisely, the
fractal particle will undergo the same number of collisions
in time $t$ as the spherical particle would undergo in time
$t^{d}$ making it a L\'evy flight with index $2/d$.

One can use the above formalism of the local fractional
Langevin equation to describe this phenomenon.
In order to do this we consider the $x$-component of the displacement
of the particle as a function of time and 
hypothesise that the frictional force is
proportional to $x^{(1/d)}(t)$, which we call the ``renormalised"  velocity 
instead of the usual velocity
which is the case when $d=1$. Here $1/d$ is
again the H\"older exponent of the $x$-component of the
fractal boundary~\cite{Mandelbrot1982}. 
A way to motivate this is to note that the friction increases
with the irregularity
of the particle and this diverging situation can
be remedied by renormalising the velocity. We use this in the Langevin
equation and again assume the high friction limit and neglect
the acceleration term. In this way we arrive at the 
equation~(\ref{eq:lengevin}) and its solution we know is
the L\'evy process with index $2/d$. It is clear that,
since $1<d<2$, the resulting process has finite mean. 

Clearly we have made many assumptions in order to understand
the essence of the problem. Firstly we have assumed a strict
mathematical boundary which is irregular down to the finest
length scales. In practice, the systems will have a lower cutoff
arising from the smoothening of the boundary at the lower
length scales or the finite density of surrounding fluid
which will give rise to a smaller number of collisions
and thus make larger flights less probable. This will lead to 
truncated L\'evy flights~\cite{Mantegna1994}
instead of the L\'evy flights. Also, other constraints, 
like the size of the system, would also force us to this
conclusion. Another assumption we have
made is that the fluid consists of point particles. The
size and shape of fluid would add to the complications 
especially if the molecules are large. This is because,
as demonstrated in~\cite{VanDamme1986} and especially
for dimension greater than 1.5, the fractal boundary
develops wiggles making some part of it inaccessible
to larger molecules. The fractal dimension of this
accessible surface, which we here call apparent dimension,
may be smaller than actual dimension and may depend 
on the shape of the approaching molecule.
It is only this apparent dimension that will be important.
Finally, we have considered a particle with only the
boundary that is fractal but this again is not essential
and one can have a porous fractal. Once again, it is the
apparent dimension which will play the role~\cite{Lewis1985}
and this dimension should be greater than the dimension
of the spherical surface in order to obtain the present
result.

Now we consider the local fractional Langevin equation
with additional noise term
\begin{eqnarray}
x^{(\alpha)}(t) = \zeta(t) + \eta(t)
\end{eqnarray}
where $\zeta(t)$ is as before the white noise and $\eta(t)$
are pulses of finite height and zero width distributed on
a random Cantor-like fractal set with dimension equal to $\alpha$. 
This could be a result of some self-organised critical
process.
Its solution is given by
\begin{eqnarray}
x(t) = L(t) + P_{\eta}(t)
\end{eqnarray}
where $L(t)$ is the L\'evy process and $P_{\eta}(t)$ is as
defined after eqn.~(\ref{eq:soln}). The second part of the solution has
log-periodic oscillations embedded at the fractal set~\cite{Kolwankar2004}.
Such log-periodic oscillations have been observed in
stockmarket data~\cite{Drozdz1999} and 
in other fields~\cite{Benkadda1997,Sornette1998}.

Finally we consider two more local stochastic differential
equations in order to demonstrate the generality of the
formalism. The first equation we consider is
\begin{eqnarray}
x^{\alpha}(t) + \eta(t) x(t) = \zeta(t)
\end{eqnarray}
and its solution is given by
\begin{eqnarray}
x(t) = x_0 e^{-P_{\eta}(t)}
+ \int_0^t e^{-(P_{\eta}(t)-P_{\eta}(t'))} \zeta(t') d^{\alpha}t'
\end{eqnarray}
with $x(0)=x_0$. And the last
equation we consider generalises the simple Kubo oscillator
\begin{eqnarray}
x^{\alpha}(t) = a \zeta(t) x(t)
\end{eqnarray}
with the solution 
\begin{eqnarray}
x(t)= x_0 \exp\left( a \int_0^t \zeta(t') d^{\alpha}t'\right)
\end{eqnarray}
This demonstrates that many different complex stochastic
signals can be generated using various local stochastic
differential equations.

To conclude, we have nontrivially and significantly expanded
the concept of stochastic differential equation formalism as
instituted by Langevin by replacing the first order derivative
with a local fractional derivative of order $\alpha$ between
zero and one. It should be emphasized that this procedure does
not add any extra correlations in the system as a nonlocal
fractional derivative would do. This naturally gives rise to L\'evy
processes as a solution when the input noise is usual
white noise putting L\'evy processes and fBm on symmetrical
dynamical footing. Essentially this result is a consequence of
the integration of the white noise over a fractal boundary and
should have wider implications. Here, as an application, we have considered the
Brownian motion of a fractal particle and argued that the
fractality of the Brownian particle would give rise to the
L\'evy flights. More precisely, a particle with fractal
boundary of dimension $d$ performs a L\'evy flight of
index $2/d$ when immersed in fluid with point particles.
In the case of fluid with larger molecules results may be
different and will have to be worked out for that case.
Furthermore we can describe this statistical effect
using the above generalised Langevin equation by hypothesising
that the frictional force is proportional to the ``renormalised"
velocity. 
This has given a new way of obtaining superdiffusive
behaviour wherein the environment or the heat bath has normal
dynamics but it is the fractality of the system that makes
it anomalous. 
Careful experiments should be carried out
in order to confirm this
prediction. 
A way to do this would be to study the Brownian motion of
an aggregate, whose dimension
is known, by tagging it with a fluorescent probe 
and measure the self-diffusion coefficient~\cite{Ott1990}.
We have restricted our attention only to the translational
motion. The behaviour of the angular velocity should also
be interesting in itself.
Clearly, there are many other factors affecting the 
Brownian motion of a real biological molecule in living
cell environment but the implications of this study 
should be taken into account.

This work also illustrates the use of local fractional calculus
to describe phenomena involving fractals. Such a tool is badly
needed for better understanding of structures and processes involving fractals
(see~\cite{Erzan1997}  for another notable approach with the same aim). 
This makes it necessary to develop this formalism further to its
full potential, especially the present generalization of the
stochastic differential equations. The corresponding extension
of the delta function might lead to a new way of characterising
the correlations.

\begin{acknowledgments}
I would like to thank J\"urgen Jost and Alok Paul for carefully reading the paper.
I am greatful to Alexander-von-Humboldt-Stiftung for 
financial
support.
\end{acknowledgments}

\bibliography{fractional}

\end{document}